\documentclass[a4paper,twoside]{article}
\usepackage{comment}
\usepackage{graphicx}
\usepackage{epsfig}
\usepackage{subfigure}
\usepackage{calc}
\usepackage{amssymb}
\usepackage{amstext}
\usepackage{amsmath}
\usepackage{amsthm}
\usepackage{multicol}
\usepackage{pslatex}
\usepackage{apalike}
\usepackage{url}
\usepackage{SCITEPRESS}     
\usepackage{algorithm}
\usepackage{algpseudocode}
\usepackage{lscape}
\usepackage{rotating}
\usepackage{lipsum}
\usepackage[UKenglish]{babel}
\usepackage{setspace}
\usepackage{color}
\newcommand*{\rom}[1]{\expandafter\@slowromancap\romannumeral #1@}
\DeclareMathOperator*{\argmax}{argmax}

\begin{document}

\title{Dynamic Pricing and Energy Management Strategy for EV Charging Stations under Uncertainties}

\author{\authorname{Chao Luo, Yih-Fang Huang, and Vijay Gupta}
\affiliation{Department of Electrical Engineering \\
University of Notre Dame, Notre Dame, Indiana, USA}
\email{\{cluo1, huang, vgupta2\}@nd.edu}
}


\abstract{This paper presents a dynamic pricing and energy management framework for electric vehicle (EV) charging service providers. To set the charging prices, the service providers faces three uncertainties: the volatility of wholesale electricity price, intermittent renewable energy generation, and spatial-temporal EV charging demand. The main objective of our work here is to help charging service providers to improve their total profits while enhancing customer satisfaction and maintaining power grid stability, taking into account those uncertainties. We employ a linear regression model to estimate the EV charging demand at each charging station, and introduce a quantitative measure for customer satisfaction. Both the greedy algorithm and the dynamic programming (DP) algorithm are employed to derive the optimal charging prices and determine how much electricity to be purchased from the wholesale market in each planning horizon. Simulation results show that DP algorithm achieves an increased profit (up to 9\%) compared to the greedy algorithm (the benchmark algorithm) under certain scenarios. Additionally, we observe that the integration of a low-cost energy storage into the system can not only improve the profit, but also smooth out the charging price fluctuation, protecting the end customers from the volatile wholesale market.}

\onecolumn \maketitle \normalsize \vfill

\noindent Symbols
\begin{itemize}
\item[] $N$: total number of planning horizon.
\item[] $s_j$: the $j$-th EV charging station.
\item[] $p_{kj}$; charging price of the $j$-th charging station in the $k$-th horizon.
\item[] $c_k$: electricity wholesale price in the $k$-th horizon.
\item[] $E$: electricity storage capital.
\item[] $R_k$: total revenue in the $k$-th horizon.
\item[] $o_k$: electricity purchase in the $k$-th horizon.
\item[] $d_{kj}$: charging demand at the $j$-th charging station in the $k$-th horizon.
\item[] $G_k$: overall customer satisfaction.
\item[] $\beta$: weighting parameter of customer satisfaction.
\item[] $\alpha$: shape parameter for customer satisfaction function.
\item[] $\omega$: shaping parameter for customer satisfaction function.
\item[] $\phi_k$: total charging demand in the $k$-th horizon.
\item[] $Q_k$: stress imposed on power grid due to EV charging in the $k$-th horizon.
\item[] $o_{\text{ref}}$: reference electricity purchase (average electricity purchase).
\item[] $o_{\textrm{max}}$: maximum electricity purchase.
\item[] $\mu$: weighting parameter of electricity purchase fluctuation.
\item[] $I_k$: electricity storage at the beginning of the $k$-th horizon.
\item[] $u_k$: renewable energy generation in the $k$-th horizon.
\item[] $W_k$: electricity storage cost in the $k$-th horizon.
\item[] $\eta$: unit electricity storage cost.
\item[] $\Pi_k$: total utility in the $k$-th horizon.
\item[] $\gamma_{i,j}$: price elasticity parameter.
\item[] $J_k(I_k)$: maximum aggregated utility from the $k$-th horizon to the last horizon.
\end{itemize}

\section{\uppercase{Introduction}}
\noindent Recent innovations in battery and powertrain technology have served as a catalyst for expediting the proliferation of electric vehicles (EVs).  EVs exhibit many advantages over the internal combustion engine (ICE) vehicles, including lower operation cost, higher fuel conversion efficiency, and reduced or eliminated tailpipe emission \cite{simpson,efficiency}. The American market share of plug-in EVs in new registered cars increased from 0.14\% to 0.37\% in 2012, 0.62\% in 2013, and 0.72\% in 2014 \cite{market}. According to Navigant Research, the global light duty EV market is expected to grow from 2.7 million vehicle sales in 2014 to 6.4 million in 2023 \cite{navigant}. EVs will play a significant role in transportation electrification. Nevertheless, the limited driving range and the long charging time are still the major obstacles to the proliferation of EVs. The ``range anxiety" is like the Sword of Damocles for EV owners. More charging stations need be established to alleviate the ``range anxiety". In addition, the profitability of the EV charging industry is another critical issue that should be considered. The EV charging industry needs a promising business model to bring more private investors into this industry instead of solely relying on financial support or incentives from governments. The effective and efficient management of charging infrastructure is at the heart of the EV charging industry. The objective of this paper is to provide guidelines for charging service providers make informed and optimized decisions on pricing and energy management so as to coherently improve profits, enhance customer satisfaction, and reduce uncertainties or risks.

Currently, there is a plethora of literature aiming at addressing the dynamic pricing issue of EV charging stations. Yan \emph{et al.} proposed a multi-tiered real-time pricing algorithm for charging stations by taking into account both the day-ahead predicted electricity price and the real-time load information \cite{yan}. However, they did not consider the possibility that EV owners may change their charging behavior in response to the varying prices. Han \emph{et al.} presented a dynamic pricing and scheduling scheme for EV charging stations while considering grid-to-vehicle (G2V) and vehicle-to-grid (V2G) \cite{han}. They used a Stackelberg game to characterize the strategic interactions between the ``selfish" EV owners and the charging stations. However, they only considered a single charging station in their model. In \cite{ban}, a price control method was employed to guide EVs to different charging stations while satisfying the predefined QoS and maintaining power grid stability. The authors used a multi-queue system to model the arrival and departure of EVs. Nevertheless, they treated the charging station as a profit-neutral entity, which may not be an appropriate assumption for the real market. In \cite{martirano}, the authors proposed a scheme called the ``Interactive Energy", for the dynamic pricing and electricity delivery of the EV charging services based on the status of the microgrid. However, the overall customer satisfaction was not considered in their analysis. The pricing models proposed in \cite{asr} and \cite{yguo} did not incorporate the renewable energy (like wind power or solar power), which is becoming an important energy source. In \cite{yguo2}, the authors addressed a two-stage framework for the economic operation of a microgrid-like electric vehicle parking deck using a stochastic approach and model predictive control (MPC).

Our work is motivated by the fact that the charging service providers face many uncertainties when determining the appropriate charging prices and managing the electricity storage. In this paper, we consider three types of uncertainties that the service providers may face: (1) the uncertainty of spatial-temporal charging demand at each charging station, (2) the uncertainty of renewable energy generation, and (3) the uncertainty of the electricity price at the wholesale market. We also assume that a charging service provider operates a network of charging stations. As a mediator in the power grid, the service provider purchases the electricity from the wholesale market and resells it to EV owners. We also assume that the service provider owns a storage system that stores the excessive electricity temporarily. Additionally, the service provider can harvest the distributed renewable energy generation, and use it as a supplementary energy source for EV charging.

In our study here, we first employ a linear regression model to estimate the EV charging demand. Specifically, the customer's price elasticity coefficients, reflecting the customer's sensitivity to charging price variation, will be estimated using historical data. Subsequently, we apply the Dynamic Programming (DP) computation algorithm to derive the optimal charging prices and how much electricity to be purchased from the wholesale market based on the current electricity storage and renewable energy forecast.

The main contribution of this paper is a computation framework to help the EV charging service provider calculate the optimal charging prices and determine the appropriate amount of electricity to purchase from the wholesale market in each planning horizon. Our computation framework can deal with the three aforementioned uncertainties and is aimed at striking a balance among the profit, customer satisfaction, and the power grid stability.

\section{\uppercase{Problem Formulation}}
\begin{figure}[htbp]
\centering
\includegraphics[width=0.52\textwidth]{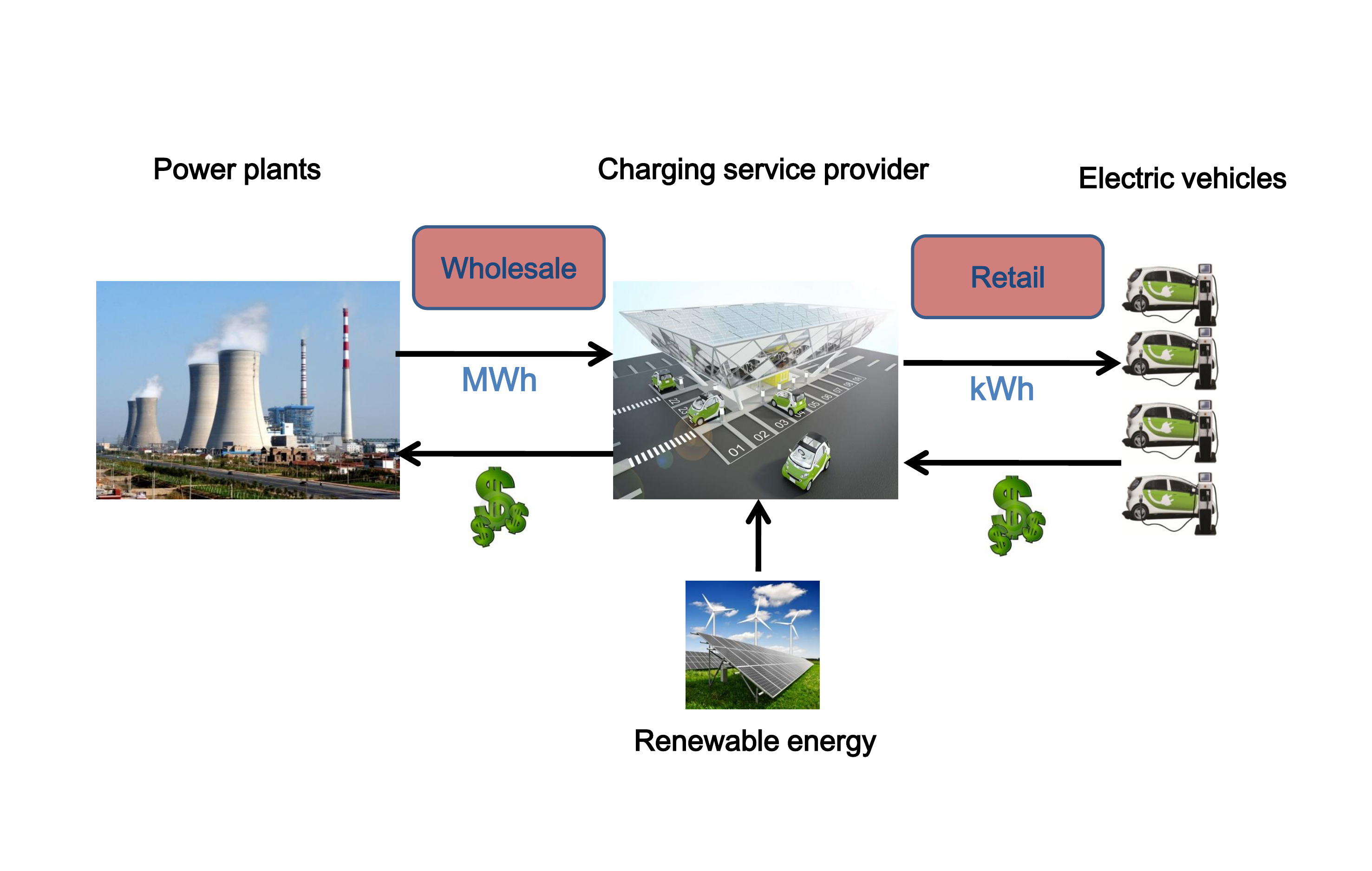}
\caption{The EV Charging Market}
\label{fig1}
\end{figure}

\noindent In our model, we postulate that there is an EV charging service provider operating a network of charging stations. As a mediator between the power grid and the end customers (i.e., EV owners), the charging service provider purchases electricity from the wholesale market at the day-ahead prices, and resells it to EV owners at the retail charging price.  Figure \ref{fig1} depicts a general business model for EV charging.

\subsection{Profit of The Service Provider}
The worldwide deregulation of electricity market (e.g., PJM Interconnection, ERCOT in USA, New Zealand, Singapore, UK markets, etc.) gives birth to the prosperous forward markets and day-ahead markets. The Independent System Operator (ISO) or the Regional Transmission Organization (RTO) calculates the day-ahead market prices through an auction between the power generators and the retailers using the locational marginal pricing (LMP) scheme \cite{huisman,lmp,lmp2}. We assume that the charging service provider is one of the retailers, buying electricity from the wholesale market and reselling it to EV owners. Let $S=[s_1,s_2,\cdots,s_L]$ denote the set of charging stations. We divide a day into $N$ planning horizons (stages). At the beginning of each horizon, the service provider updates the charging prices, and calculates how much electricity needs to be purchased from the wholesale market. We allow charging prices vary among different charging stations. Let $P_k=[p_{k1},p_{k2},\cdots,p_{kL}](k=1,2,\cdots,N)$ be the charging price vector in the $k$-th horizon, and $o_k$ be the electricity purchase. Currently, the day-ahead market prices are calculated on a hourly basis, so $N=24$. Let $C=[c_1,c_2,\cdots,c_N]$ denote the day-ahead wholesale market prices. The total profit of the service provider in the $k$-th horizon is given by

\begin{equation}
R_k=\sum_{j=1}^Lp_{kj}d_{kj}-c_ko_k\;\;(k=1,2,\cdots,N),
\end{equation}
where $d_{kj}$ is the charging demand at the $j$-th charging station in the $k$-th horizon, and $\sum_{j=1}^Lp_{kj}d_{kj}$ is the total revenue, and $c_ko_k$ is the cost of electricity purchased in the $k$-th horizon.

\subsection{Customer Satisfaction Evaluation}
 The charging service provider attempts to achieve the goals of improving the profits, enhancing the customer satisfaction, and maintaining power grid stability. Poor customer satisfaction may hinder the wide adoption of EVs, thus, affecting the development of the entire EV industry. In this sense, the charging service provider cannot be a myopic profit squeezer that maximizes the profit at the expense of customer satisfaction. Various customer satisfaction evaluation methods have been studied in \cite{yang,fahrioglu,faranda}. In this paper, we use a quadratic function to characterize the overall customer satisfaction of the entire population of EV owners, denoted by $G_k$.

 \begin{equation}
\label{satisfactioneq}
G_k=\beta\left(\omega\phi_k-\frac{\alpha}{2}\phi_k^2\right),\;0\leq \phi_k \leq E
\end{equation}
where $\beta$ is the weighting parameter and $E$ is the capacity of the electricity storage system, $\omega$ and $\alpha$ are the shape parameters of this quadratic function. The variable $\phi_k$ is the total electricity consumption (charging demand) of all EV owners in the $k$-th horizon which is defined as

\begin{equation}
\phi_k=\sum_{j=1}^Ld_{kj}.
\end{equation}

\begin{figure}[htp]
  \centering
  \includegraphics[width=3.5in]{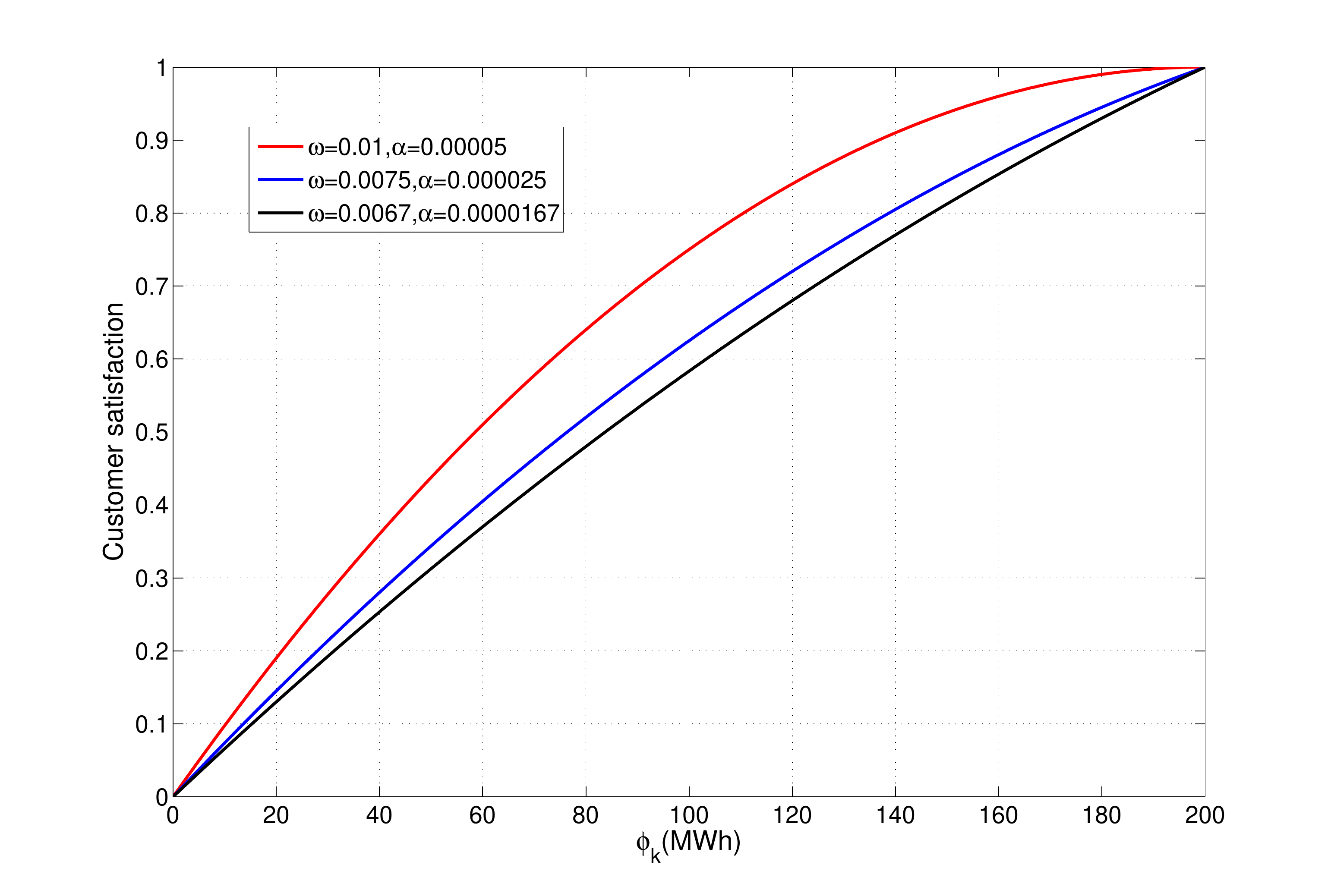}
  \caption{Customer Satisfaction Functions ($E=200$)}\label{satisfaction}
\end{figure}

The quadratic functions with different combinations of $\omega$ and $\alpha$ are shown in Figure \ref{satisfaction}. We observe that the quadratic function has a minimum value of 0, suggesting that the EV owners are very ``unhappy", and a maximum value of 1, suggesting that the EV owners are very ``happy". Additionally, Equation (\ref{satisfactioneq}) is a non-decreasing concave function with a non-increasing first-order derivative. The overall customer satisfaction grows as the total electricity consumption increases. However, the decreasing growth rate suggests that the customer satisfaction tends to get saturated as the electricity consumption increases.

\subsection{Impact of EV Charging on Power Grid}
Many studies \cite{lopes,kinter,scott} have shown that large-scale simultaneous EV charging presents many challenges to the existing power grid pertaining to severe power loss, power grid stability, frequency drift, and voltage fluctuation, etc. In the electric power system, the networked generators cooperatively adjust their outputs to balance the supply and the demand and maintain the power quality. Generally, the power generators hope that the load is predictable and relatively stable (or at least slow-varying). If the load fluctuates too much, the power generators have to ramp up and down frequently, resulting in low efficiency and high maintenance cost. As a result, we do not want the electricity purchase from the wholesale market $o_k$ to fluctuate too much which may create a heavy ``burden" on the power grid. We formulate the penalty of EV charging in the following way
\begin{equation}
Q_k=\mu(o_k-o_{\textrm{ref}})^2,
\end{equation}
where $o_{\textrm{ref}}$ is a reference purchase (or average electricity purchase) and $o_k$ is the electricity purchased in the $k$-th horizon. The variable $\mu$ is the weighting parameter reflecting the sensitivity of electricity purchase fluctuation.

\subsection{Cost of Electricity Storage}
We assume that the charging service provider has an electricity storage with a capacity of $E (\textrm{MWh})$. Let $I_k$ denote the electricity in the storage at the beginning of the $k$-th horizon, and let $u_k$ be the renewable energy generation (i.e. wind power or solar power). Here $u_k$ is the predicted renewable energy. The electricity storage cost in the $k$-th horizon is given as follows
\begin{equation}
W_k=\eta(I_k+u_k+o_k-\sum_{j=1}^Ld_{kj}),
\end{equation}
where $\sum_{j=1}^Ld_{kj}$ is the total charging demand in the $k$-th horizon, and $\eta(\$/\textrm{MWh})$ is the unit storage cost. The storage cost includes capital cost, maintenance cost, and power loss due to energy conversion.

Finally, the total utility of the service provider in the $k$-th horizon is given as
\begin{equation}
\begin{aligned}
\Pi_k &= R_k+G_k-Q_k-W_k\\
&=\sum_{j=1}^Lp_{kj}d_{kj}-c_ko_k+\beta\left(\omega\phi_k-\frac{\alpha}{2}\phi_k^2\right)-\\
&\mu(o_k-o_{\textrm{ref}})^2-\eta(I_k+u_k+o_k-\sum_{j=1}^Ld_{kj}),\\
\end{aligned}
\end{equation}

Note that the total utility consists of four components---total revenue, customer satisfaction, power grid influence, and electricity storage cost. The values of $\beta$ and $\mu$ reflect the weights of customer satisfaction and EV charging penalty in the total utility function.

Our objective here is to maximize the overall utility by solving the following optimization problem.

\begin{equation}\label{eq1}
\begin{aligned}
&(P_1^*,o_1^*,\cdots,P_N^*,o_N^*)=\argmax_{P_1,o_1,\cdots,P_N,o_N}
\left\{\sum_{k=1}^N\Pi_k\right\},\\
&\;\;\; s.t.\\
&\begin{cases}
0\leq o_k\leq o_{\textrm{max}};k=1,2,\cdots,N\\
p_{kj}\geq0;j=1,2,\cdots,L\\
 I_k + o_k-\sum_{j=1}^Ld_{kj}\geq0\\
I_k + o_k-\sum_{j=1}^Ld_{kj}\leq E\\
d_{kj}\geq0;j=1,2,\cdots,L
\end{cases}
\end{aligned}
\end{equation}
where $P_k$ and $o_k$ are , respectively, the charging price vector and electricity purchase in the $k$-th horizon.

To resolve this problem, we are facing two major challenges: (1) accurately estimate the charging demand $\phi_k$, and (2) solve the large-scale optimization problem in a more efficient way. In the following sections, we will discuss these problems in more details.

\section{\uppercase{Spatial-temporal Charging Demand Estimation}}
\noindent In this section, we consider the estimation of the charging demand $\phi_k$. The charging demand function characterizes the customer's responsiveness to the fluctuation of charging prices, for EV owners may adjust their charging demand or schedule in response to the variation of charging prices.

In the initial phase of the optimization framework, we do not have the information of the EV owner's responsiveness to the charging prices. We thus apply a linear regression model to learn and predict the charging demand $\phi_k$. For each charging station $j(j=1,2,\cdots,L)$, the charging demand is expressed by

\begin{equation}\label{lrm}
\begin{cases}
d_{k1}=\gamma_{0,1}-\gamma_{1,1}p_{k1}+\gamma_{2,1}p_{k2}+\cdots+\gamma_{N,1}p_{kN}+\epsilon_{k1},\\
d_{k2}=\gamma_{0,2}+\gamma_{1,2}p_{k1}-\gamma_{2,2}p_{k2}+\cdots+\gamma_{N,2}p_{kN}+\epsilon_{k2},\\
\vdots\\
d_{kL}=\gamma_{0,L}+\gamma_{1,L}p_{k1}+\gamma_{2,2}p_{k2}+\cdots-\gamma_{N,L}p_{kN}+\epsilon_{kL},\\
\end{cases}
\end{equation}
where $\gamma_{0,j}(j=1,2,\cdots, L)$ is the intercept of the $j$-th linear regression equation, and $\gamma_{i,j}=\gamma_{j,i}(i\neq j)$ are the cross-price elasticity parameters, reflecting how the change of the charging price of station $j$ can influence the charging demand at station $i$. And $\gamma_{i,i}$ is the self-price elasticity parameter, reflecting how the change of the charging price of station $i$ can influence its own charging demand.

In this work, we employ the recursive least square (RLS) \cite{proakis} method to estimate the elasticity demand parameters from historical data. Let $W_j=[\gamma_{0,j},\gamma_{1,j},\cdots,\gamma_{N,j}]$ denote the price elasticity parameter vector relevant to charging station $j(j=1,2,\cdots,L)$. Applying the RLS algorithm, we have the following update formula

\begin{equation}
\begin{aligned}
\begin{cases}
&e_{kj}=d_{kj}-P_k^{\textrm{T}}W_j,\\
&g_{kj}=\frac{H_{(k-1)j}P_k}{\lambda+P_{k}^{\textrm{T}}H_{(k-1)j}P_k},\\
&H_{kj}=\lambda^{-1}H_{(k-1)j}-g_{kj}P_k^{\textrm{T}}\lambda^{-1}P_k,\\
&W_j\leftarrow W_j+e_{kj}g_{kj},\\
\end{cases}
\end{aligned}
\end{equation}
where $e_{kj}$ is the prediction error and $\lambda$ is the forgetting factor. In initialization, $H_{0j}$ is the identity matrix and $P_0$ is an all-zero vector.

Note that Equation (\ref{lrm}) captures both the spatial and temporal fluctuation of charging demand. The difference in population density, traffic flow, and urbanization level may result in the spatial fluctuation of charging demand. Thus, we use different linear regression equations to estimate different charging stations. On the other hand, the use of RLS algorithm enables us to characterize the temporal fluctuation of charging demand. It keeps track of the most recent changes in customer's charging behavior because the price elasticity parameters will be updated once a new data is observed.

\section{\uppercase{Pricing Policies: Greedy Algorithm vs DP Algorithm}}
\noindent Note that Equation (\ref{eq1}) is a complex optimization problem with $N(L+1)$ decision variables and $N(2L+3)$ constraints. It is mathematically cumbersome and hardly feasible to solve this problem in a brute force manner. One approach is to divide the original optimization problem into $N$ independent subproblems. Each horizon corresponds to a subproblem, and then employ the greedy
search algorithm. This idea will be further discussed in Subsection 4.1.

On the other hand, we observe that the original problem exhibits the properties of overlapping subproblems and optimal substructure, which can be exploited to solve this problem more efficiently. Here, we apply the dynamic programming (DP) computation algorithm to the original problem. DP is a computation algorithm of solving a large-scale complex problem by partitioning it into a set of smaller and simpler subproblems \cite{cormen,bertsekas}. By solving and combining these subproblems in a forward (bottom-up) or backward (top-down) fashion, we can obtain the solution to the original problem. In contrast to the brute force approach, DP can significantly accelerate computation speed and save storage. We will discuss DP in Subsection 4.2.

\subsection{Greedy Algorithm}
The original optimization problem in Equation (\ref{eq1}) aims to maximize the total utility over $N$ horizons. The control variables are ``chained" in the sense that the decision variables in the previous horizon can influence the decision variables in the current horizon. For simplicity, we ignore the correlation between adjacent horizons, and try to maximize the utility in each individual horizon. Specifically, we attempt to solve the following problem in the $k$-th horizon,

\begin{equation}\label{greedy}
\begin{aligned}
&(P_k^*,o_k^*)=\argmax_{P_k,o_k}\left\{\Pi_k \right\}\\
&\;\;\; s.t.\\
&\begin{cases}
0\leq o_k\leq o_{\textrm{max}};k=1,2,\cdots,N\\
p_{kj}\geq0;j=1,2,\cdots,L\\
 I_k + o_k-\sum_{j=1}^Ld_{kj}\geq0\\
I_k + o_k-\sum_{j=1}^Ld_{kj}\leq E\\
d_{kj}\geq0;j=1,2,\cdots,L
\end{cases}
\end{aligned}
\end{equation}
where $P_k$ and $o_k$ are , respectively, the charging price vector and electricity purchase in the $k$-th horizon. We will use the  greedy algorithm as a benchmark in the simulations.

\subsection{Dynamic Programming Algorithm}
Note that the hourly-based wholesale electricity prices are only posted day ahead. We analyse the dynamic pricing problem with finite horizons (stages) with $N=24$. The system dynamics are expressed by the evolution of some variables, or the system's state variables, under the influence of the decision variables at the beginning of each horizon (stage) \cite{bertsekas,nemhauser}. The system dynamics are expressed by the following evolution equation
\begin{equation}
\begin{aligned}
I_{k+1}&=I_{k}+u_k+o_k-\phi_k\\
&=I_{k}+u_k+o_k-\sum_{j=1}^Ld_{kj}\\
&=f(I_k,u_k,P_k,o_k),k=1,2,\cdots,N\\
\end{aligned}
\end{equation}
where $I_k$ is the state variable, representing the electricity storage at the beginning of the $k$-th horizon. The variables $u_k$ and $o_k$ are, respectively, the renewable energy and the electricity to be purchased from the wholesale market. The charging demand in the $k$-th horizon is $\phi_k$. Note that $\phi_k$ is actually a function of the charging price vector $P_k$, and the decision variables of the system are $(P_k, o_k)$. The aggregated utility of the service provider from the first horizon to the $N$th horizon is given by

\begin{equation}
\Pi_{N+1}(I_{N+1})+\sum_{k=1}^N\Pi_k(I_k,P_k,o_k),
\end{equation}
where $\Pi_{N+1}(I_{N+1})$ is the terminal utility incurred at the end of the process. We can assign a heuristic value for the terminal utility. The maximum utility $J_1(I_1)$  is  given by the following form

\begin{equation}\label{eq2}
\begin{aligned}
&J_1(I_1) =\\ &\max_{P_1,o_1,\cdots,P_N,o_N}\left\{\Pi_{N+1}(I_{N+1})+\sum_{k=1}^N\Pi_k(I_k,P_k,o_k)\right\},
\end{aligned}
\end{equation}
Furthermore, the utility $J_1(I_1)$ can be calculated in a recursive manner as follows

\begin{equation}\label{eq3}
J_1(I_1)=\max_{P_1,o_1}\left\{\Pi_1(I_1,P_1,o_1)+J_2(I_2)\right\},
\end{equation}
or
\begin{equation}\label{eq3}
J_1(I_1)=\max_{P_1,o_1}\left\{\Pi_1(I_1,P_1,o_1)+J_2(f(I_1,u_1,P_1,o_1))\right\},
\end{equation}
where $J_2(I_2)$ is given by

\begin{equation}
\begin{aligned}
&J_2(I_2) =\\ &\max_{P_2,o_2,\cdots,P_N,o_N}\left\{\Pi_{N+1}(I_{N+1})+\sum_{k=2}^N\Pi_k(I_k,P_k,o_k)\right\},
\end{aligned}
\end{equation}
We can apply Equation (\ref{eq3}) recursively from the $N$th horizon backward to the first horizon to derive the solution $J_1(I_1)$. The detailed derivation of Equation (\ref{eq2}) to Equation (\ref{eq3}) is given in Appendix.

Let $X_k=[p_{k1},p_{k2},\cdots, p_{kL},o_k]^{\textrm{T}}$ denote the decision variables. Moreover, the recursive DP formula can be rewritten as one of quadratic programming as follows,

\begin{equation}\label{quad}
J_k(I_k)=\max_{X_k\in Z(X_k)}\Big\{\frac{1}{2}X_k^{\textrm{T}}\mathbf{Q}X_k+B^{\textrm{T}}_kX_k+ r_k\Big\},
\end{equation}
where $Z(X_k)$ is the feasible solutions derived from the constraints in Equation (\ref{eq1}).
\noindent The matrix $\mathbf{Q}$ is given by
\begin{equation}
\left[
    \begin{array}{cccc}
    -2\gamma_{1,1} - \alpha\beta\Gamma_1^2 & \cdots & 2\gamma_{1,L} - \alpha\beta\Gamma_1\Gamma_L&0\\
    2\gamma_{2,1}-\alpha\beta\Gamma_2\Gamma_1 & \cdots & 2\gamma_{2, L} -\alpha\beta\Gamma_2\Gamma_L& 0 \\
    \vdots & & & \vdots\\
    2\gamma_{L,1} - \alpha\beta\Gamma_L\Gamma_1& \cdots & -2\gamma_{L,L} - \alpha\beta\Gamma_L^2& 0\\
    0 & \cdots & 0 & -\mu\\
    \end{array}
\right]
\end{equation}
where $\Gamma_j(j = 1, 2,\cdots,L)$ is

\begin{equation}
\Gamma_j = -\gamma_{j,j} + \sum_{i = 1,i\neq j}^L\gamma_{j,i}.
\end{equation}

$B_k$ is
\begin{equation}
B_k =
\left[
    \begin{array}{c}
    \gamma_{0,1}+(\eta+\beta\omega)\sum_{j=1}^N\gamma_{1,j}-\alpha\beta\Gamma_0\Gamma_1\\
    \vdots\\
    \gamma_{0,L}+(\eta+\beta\omega)\sum_{j=1}^N\gamma_{L,j}-\alpha\beta\Gamma_0\Gamma_L\\
    -c_k-\eta+2\mu o_{\textrm{ref}}
    \end{array}
\right],
\end{equation}
where $\Gamma_0$ is

\begin{equation}
\Gamma_0=\sum_{i=1}^L\gamma_{0,i}.
\end{equation}

$r_k$ is

\begin{equation}
\begin{aligned}
r_k=&-\eta(I_k+u_k)+\beta\phi_k-\mu o_{\textrm{ref}}^2+\\
&(\eta+\beta\omega)\Gamma_0-\frac{\alpha\beta}{2}\Gamma_0^2+J_{k+1}(I_{k+1}),
\end{aligned}
\end{equation}
where $J_{k+1}(I_{k+1})$ is the total aggregated utility starting from the ($k+1$)th horizon to the $N$th horizon, which can be calculated using the DP recursive formula. We can treat $J_{k+1}(I_{k+1})$ as a constant value when we calculate $J_k(I_k)$.

\section{\uppercase{Dynamic Pricing and Energy Management Framework Summary}}
\begin{figure}[htbp]
\centering
\includegraphics[width=0.45\textwidth]{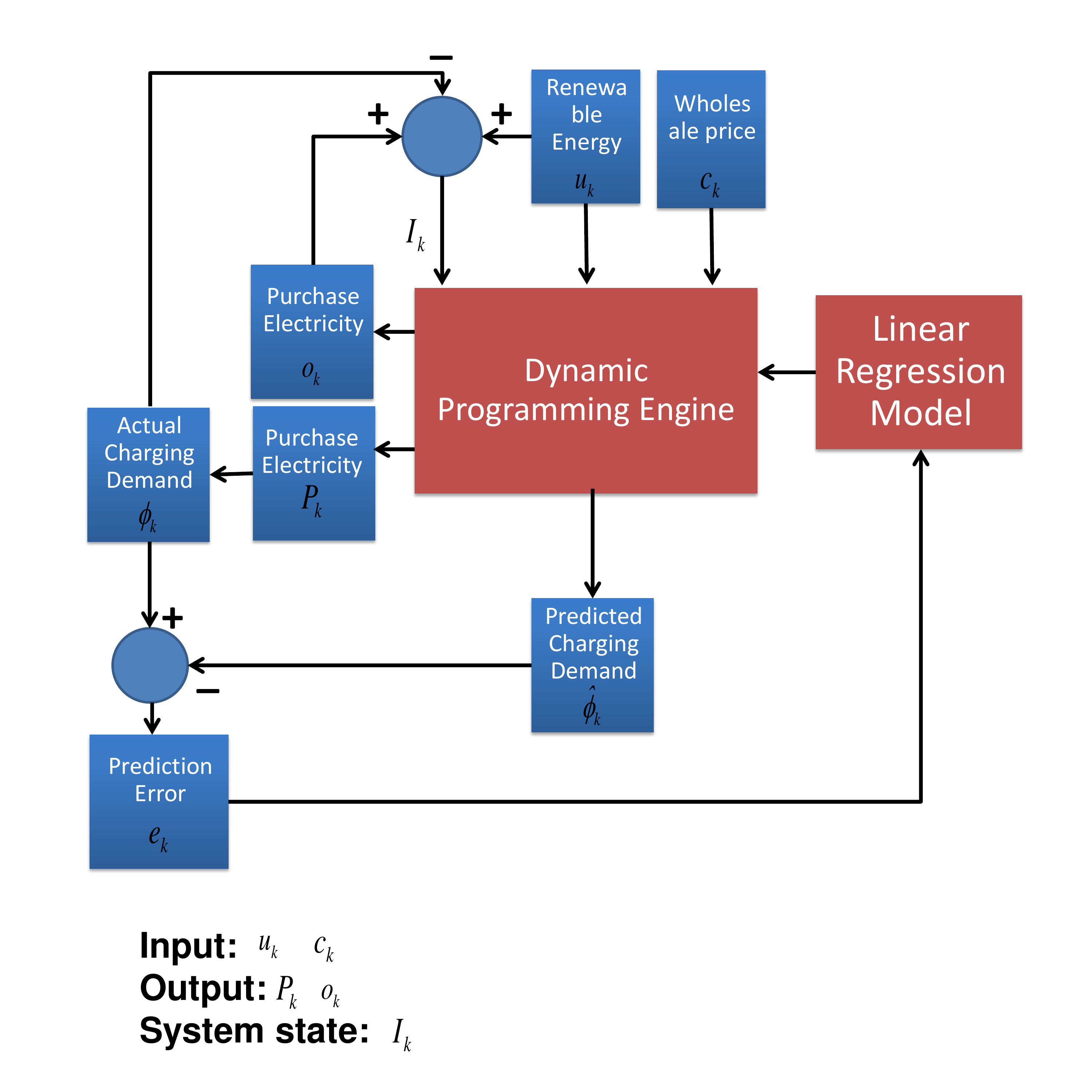}
\caption{Dynamic Pricing and Energy Management Algorithm}
\label{fig2}
\end{figure}

\noindent There are two principal modules in the dynamic pricing and energy management framework: the charging demand prediction module and the DP module. Figure \ref{fig2} illustrates the schematics of the framework. They work collaboratively to make the optimal decisions on charging prices $P_k$ and the electricity purchase $o_k$ for the service provider. The algorithm is summarized below:

\begin{algorithm}[htbp]
\caption{ Dynamic Pricing and Energy Management}
\begin{algorithmic}[1]
\Require
\\
The electricity storage (system state), $I_k$;\\
The renewable energy prediction $u_k,u_{k+1},\cdots,u_N$;\\
The wholesale electricity prices, $c_k,c_{k+1},\cdots,c_N$;
\Ensure
The new system state $I_{k+1}$, the charging prices $P_k$, and electricity purchase $o_k$;
\State Load the price coefficients $\gamma_{i,j}^k(i, j=1,2,\cdots,N)$ from linear regression module into the DP engine module;
\State The DP engine takes the inputs and generates the outputs $P_k,o_k$ using Eq. (\ref{quad});
\State Compute the charging demand prediction error $e_k=\phi_k-\hat{\phi}_k$. Apply the RLS method to update the price coefficients $\gamma_{i,j}^{k+1}=f(\gamma_{i,j}^k,e_k)$;
\State Update the electricity storage $I_{k+1} = I_k+u_k+o_k-\phi_k$;
\Return $I_{k+1},P_k,o_k$;
\end{algorithmic}
\end{algorithm}

\section{\uppercase{Simulation Results and Discussions}}
\begin{figure*}[htbp]
  \begin{minipage}[b]{0.5\linewidth}
    \includegraphics[width=3.3in]{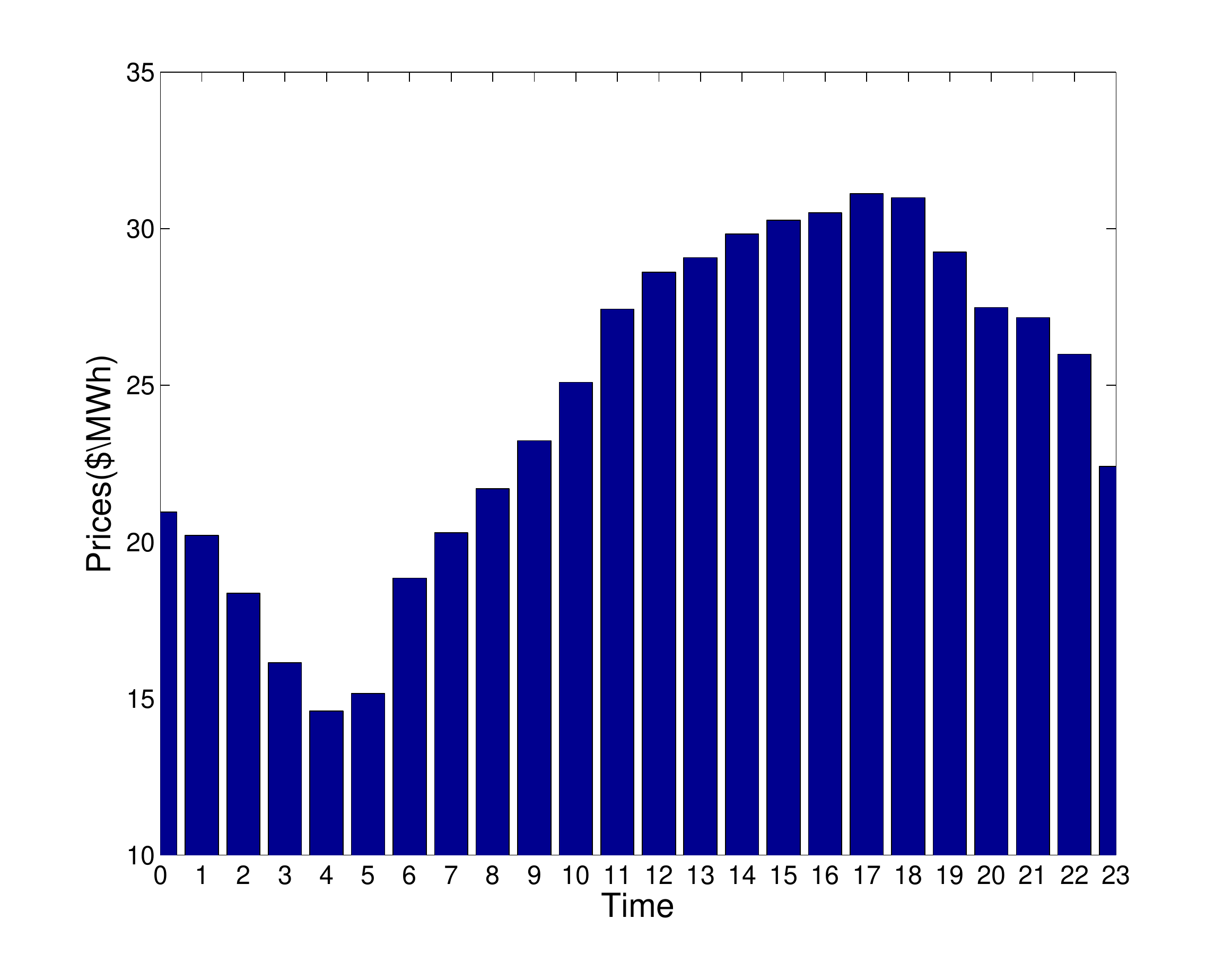}
    \caption{PJM Electricity Wholesale Prices}
    \label{fig3}
  \end{minipage}
  \begin{minipage}[b]{0.5\linewidth}
    \includegraphics[width=3.3in]{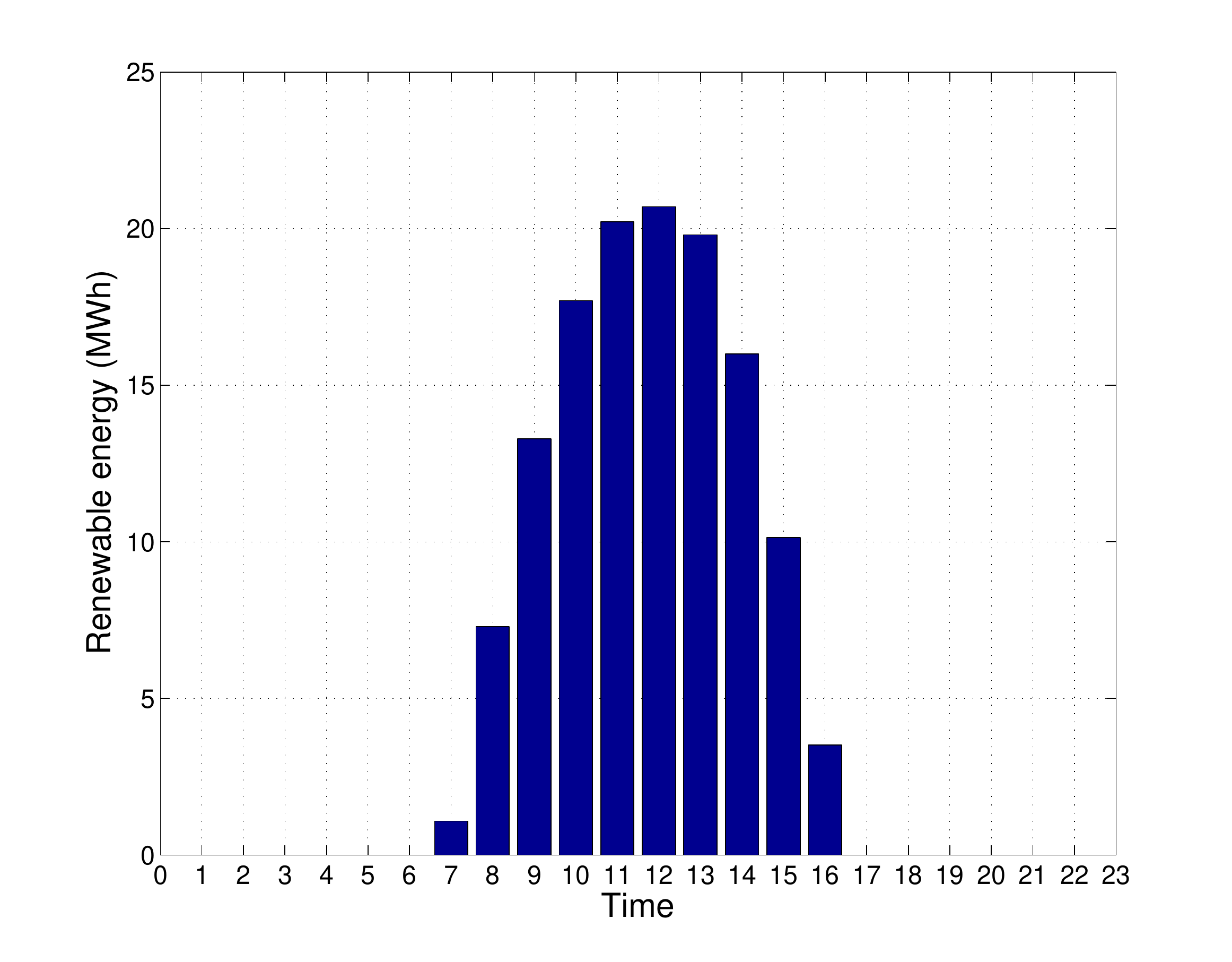}
    \caption{Hourly Renewable Energy Generation}
    \label{fig4}
  \end{minipage}
  \end{figure*}

\noindent The simulation parameters are given in Table 1. We use the historical data of the PJM day-ahead market in our simulations, see Figure \ref{fig3}. We use the solar radiation data from the National Solar Radiation Data Base \cite{solardata} as a proxy of the predicted renewable energy generation. For simplicity, we assume that the solar cell efficiency is 20\%. The renewable energy generation prediction is shown in Figure \ref{fig4}. We notice that the solar power generation begins at 8:00 and ends at 17:00 with a peak at 13:00.

\begin{table*}[!htbp]\label{ta1}
\caption{Simulation Parameters}
\begin{center}
\begin{tabular}{llll}
\hline
Coefficient & Description & Unit & Value\\

$N$ & Number of horizons & - &24\\

$E$ & Energy storage capacity & MWh & 200\\

$\omega$ & Customer satisfaction para. & - & 0.01\\

$\alpha$ & Customer satisfaction para. & - & 5e-5\\

$\beta$ & Satisfaction Parameter& - & 0, $\cdots$, 30000\\

$\mu$ & Power grid impact parameter & - & 0.1\\

$\eta$ & Storage cost  & \$/MWh &0.5, 1.0, 1.5\\

$o_{\textrm{ref}}$ & Reference purchase & MWh & 40\\
\hline
\end{tabular}
\end{center}
\end{table*}

\subsection{DP Algorithm versus Greedy Algorithm}
We use the greedy algorithm as the benchmark, and compare DP algorithm with the greedy algorithm. Figure \ref{fig5} shows the profit increase of DP algorithm (using greedy algorithm as the benchmark).  The simulation reveals that DP algorithm achieves up to 9\% increase in profit in contrast to the greedy algorithm. The reason why DP algorithm can achieve a higher profit is that it exploits the information of the entire hourly day-ahead prices and the renewable energy prediction to make optimized decisions at each horizon. The decisions made in each horizon are optimized so that the aggregated profit over multiple horizons is maximized. In contrast, the greedy algorithm is a myopic algorithm because it only maximizes the profit in the current horizon without considering the day-ahead prices and the renewable energy generation in the future. Comparing the computational complexity of the two algorithms, we note that greedy algorithm has a linear time complexity $O(N)$, while DP algorithm has a quadratic time complexity $O(N^2)$, where $N$ is the number of planning horizons. Therefore, DP algorithm achieves a higher profit (better performance) at the cost of increased computing time.

\subsection{Tradoff between Profit And Customer Satisfaction}
\begin{figure*}
  \begin{minipage}[b]{0.5\linewidth}
    \includegraphics[width=3.3in]{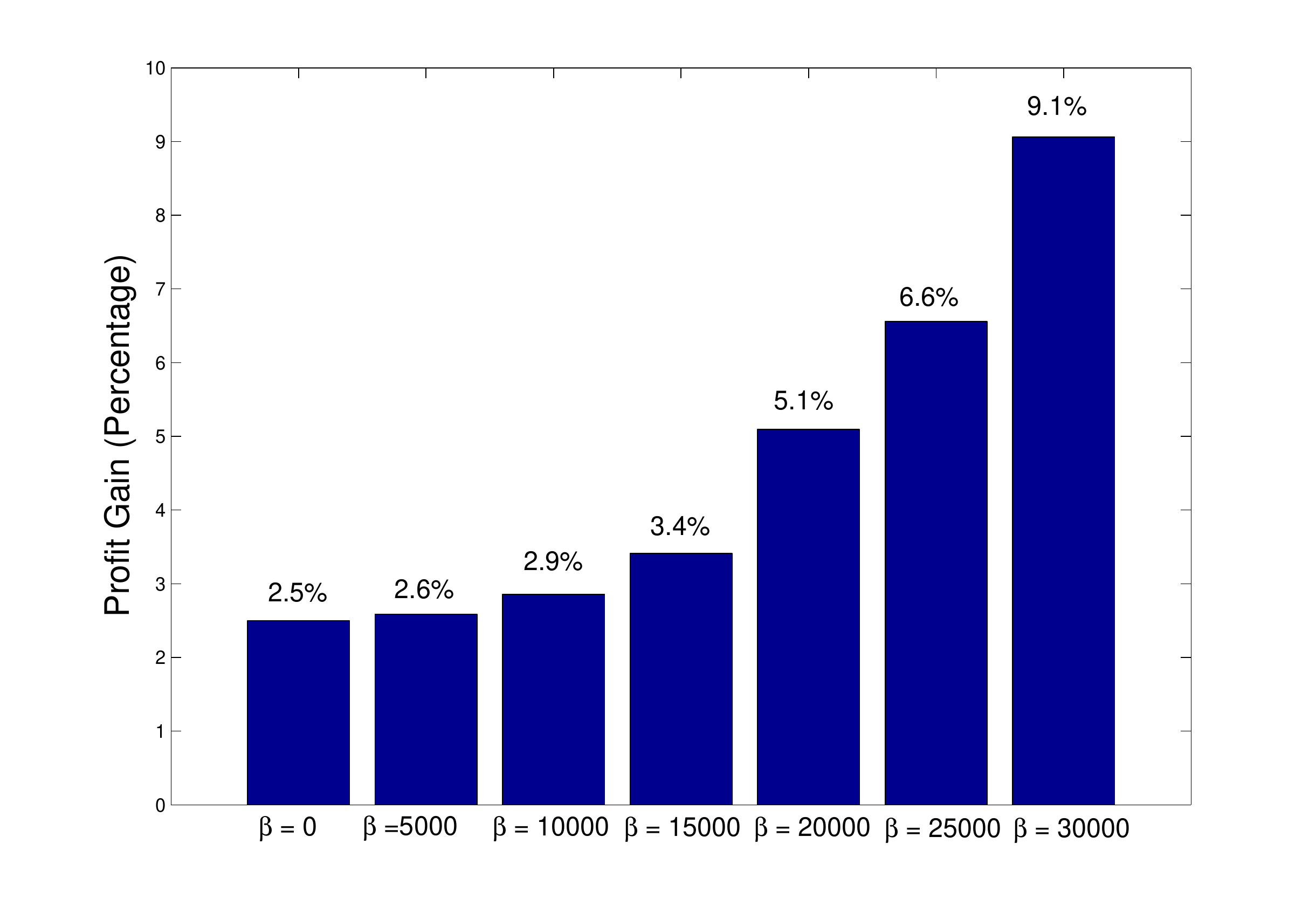}
    \caption{DP Profit Increase Percentage}
    \label{fig5}
  \end{minipage}
  \hfill
  \begin{minipage}[b]{0.5\linewidth}
    \includegraphics[width=3.3in]{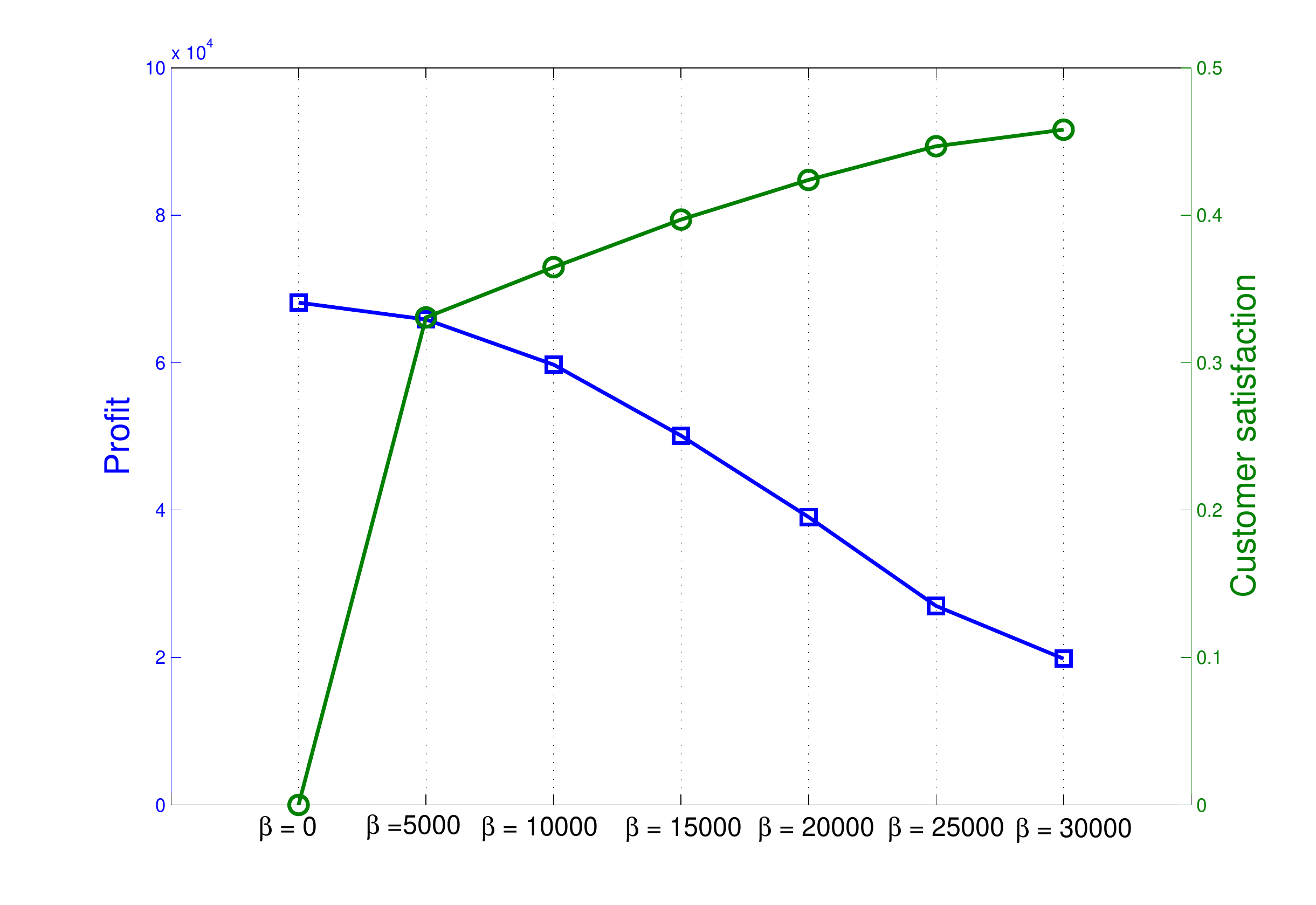}
    \caption{Total Profit vs Customer Satisfaction}
    \label{fig6}
  \end{minipage}
\end{figure*}

This section considers how the profit and customer satisfaction change as the customer satisfaction weighting parameter $\beta$ increases from 0 to 30000 with an interval of 5000. From Figure \ref{fig6}, we observe that as $\beta$ increases, the customer satisfaction increases and the profit suffers a significant decrease. It is clear that the charging service provider should make a tradeoff between profit maximization and customer satisfaction improvement by choosing a proper weighting parameter $\beta$.

\begin{figure*}[htbp]
  \begin{minipage}[b]{0.5\linewidth}
    \includegraphics[width=3.3in]{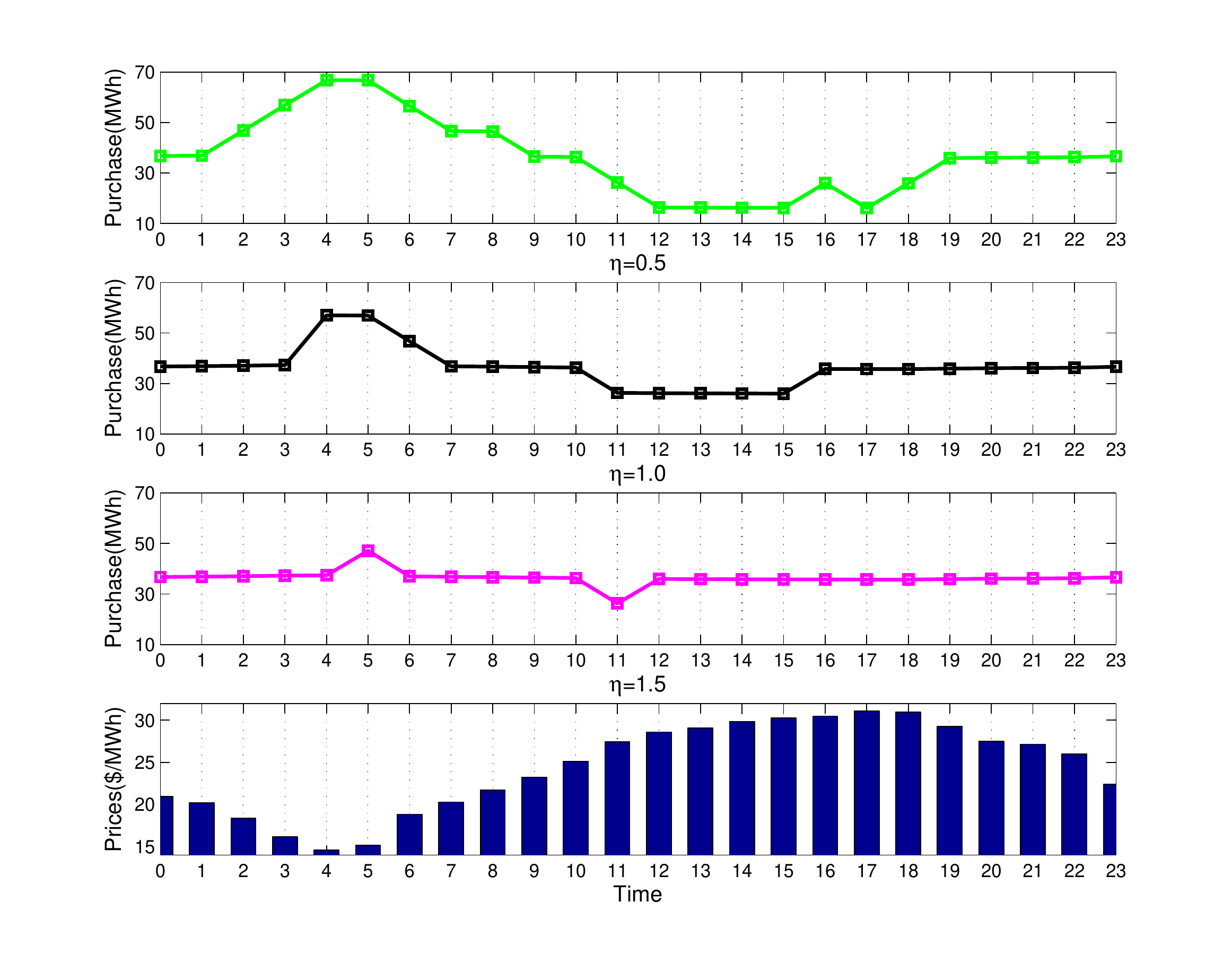}
    \caption{Electricity Purchase with Different Storage Cost}
    \label{fig7}
  \end{minipage}
  \hfill
  \begin{minipage}[b]{0.5\linewidth}
    \includegraphics[width=3.3in]{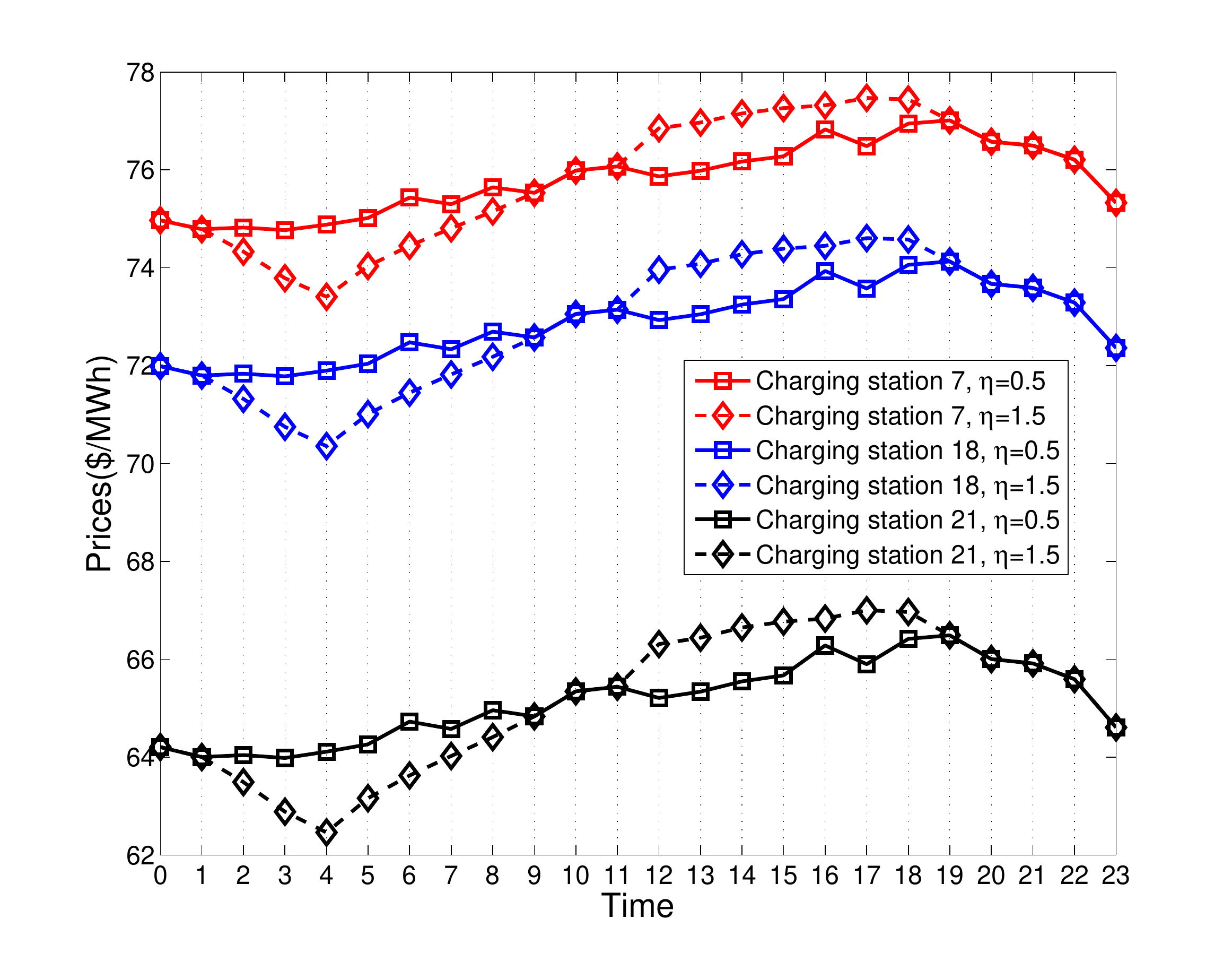}
    \caption{Charging Prices with Different Storage Cost}
    \label{fig8}
  \end{minipage}
\end{figure*}

\subsection{The Aggressive or Conservative Electricity Purchase Strategy}
The electricity storage system enables the charging service provider to purchase extra electricity from the wholesale market when the wholesale price is low, and store the unsold electricity for future use when the wholesale price is high. In this simulation, we analyze how this ``buy low and sell high" strategy may change as the energy storage cost increases. In Figure \ref{fig7}, the first three subplots are the electricity purchase with different energy storage costs ($\eta = 0.5,\eta = 1.0, \eta = 1.5$ ), and the last subplot is the day-ahead wholesale market prices. From Figure \ref{fig7}, we make three observations: (1) The average electricity purchase is 37 MWh in each horizon; (2) When $\eta = 1.5$, the electricity purchase almost does not change. This suggests that the service provider becomes conservative in electricity purchase as the storage cost increase. In other words, the service provider cannot improve the profit through ``buy low and sell high" strategy due to the high storage cost; (3) When $\eta = 0.5$ and $\eta = 1.0$, the service provider is likely to purchase more electricity during low-price horizons (from 3:00 to 8:00), and purchase less electricity during high-price horizons (from 11:00 to 19:00). Generally speaking, low electricity storage cost spurs the service provider to adopt an aggressive electricity purchase strategy.

\subsection{Smoothing Price Fluctuation via Electricity Storage System}
In this section, we investigate the correlation between the charging prices and the electricity storage cost. In the simulation we have 20 charging stations in total, and we randomly choose 3 charging stations to plot Figure \ref{fig8}. The solid lines represent the charging prices with low storage cost ($\eta = 0.5$), and the dash lines are the charging prices with high storage cost ($\eta = 1.5$).

First, we notice that different charging stations have different charging prices. Second, the charging prices with high storage cost are more volatile than those with low storage cost. When the wholesale prices are low (from 1:00 to 8:00), the charging prices with high storage cost are lower than those with low storage cost. When the wholesale prices are high (from 12:00 to 19:00), the charging prices with high storage cost are higher than those with low storage cost. The reason for the difference is that when the storage cost is low, the service provider can have more electricity reserved in the storage system which can be used in the future when the wholesale electricity price is high. Therefore, the charging prices stay relatively stable over time. As the storage cost increases, electricity storage becomes expensive. Without the ``buffer effect" of the electricity storage system, the EV owners are exposed to the varying charging price which is directly influenced by the wholesale market. Hence, a low-cost energy storage system can not only increase the total profit but also act as a buffer to smooth out the fluctuation of the charging prices.

\section{\uppercase{Conclusion}}
\noindent In this paper, a DP based pricing and energy management framework for EV charging stations is studied. The proposed framework aims to strike a balance among three conflicting goals of improving the total profit, enhancing the user satisfaction, and reducing the EV charging impact on the power grid. In this study here, we incorporate the electricity storage system and the renewable energy generation as an energy supplement. To solve the optimization problem, we apply the DP algorithm to calculate the charging prices and the electricity purchase for each planning horizon. The simulation results show that the DP algorithm can obtain higher profits compared with the greedy algorithm. In addition, we observe that the electricity purchase is heavily influenced by the wholesale prices and the energy storage cost. A low-cost energy storage system is beneficial for improving the profit and stabilizing the charging prices.

\vfill
\bibliographystyle{apalike}
{\small
\bibliography{example}}

\section*{\uppercase{Appendix}}
\noindent Given the following optimization problem
\begin{equation}\label{eq4}
\begin{aligned}
&J_k(I_k)=\\
&\max_{P_k,o_k,\cdots,P_N,o_N}\left\{\Pi_{N+1}(I_{N+1})+\sum_{j=k}^N\Pi_j(I_j,P_j,o_j)\right\}.
\end{aligned}
\end{equation}
Note that $I_{j+1}=I_{j}+u_j+o_j-\phi_j;(j>k)$, so $I_{j+1}$ is a function of $I_{j}, o_j, \textrm{and } P_j$. We can prove that $I_{j+1}$ is actually a function of $(I_k,o_k,o_{k+1},o_j,P_k,\cdots,P_j)$ by recursively applying the formula to substitute $I_j$. Then we can rewrite Equation (\ref{eq4}) as follows,
\begin{equation}
\begin{aligned}
J_k(I_k)&=\max_{P_k,o_k,\cdots,P_N,o_N}\bigg\{\Pi_{N+1}(I_{N+1})+\\
&\;\;\;\sum_{j=k}^N\Pi_j(I_k,P_k,\cdots,P_{j},o_k,\cdots,o_j)\bigg\}\\
&=\max_{P_k,o_k}\Bigg\{\Pi_k(I_k,P_k,o_k)+\\
&\;\;\;\;\max_{P_{k+1},o_{k+1},\cdots,P_N,o_N}\bigg\{\Pi_{N+1}+\sum_{j=k+1}^N\Pi_j\bigg\}\Bigg\}\\
&=\max_{P_k,o_k}\big\{\Pi_k(I_k,P_k,o_k)+J_{k+1}(I_{k+1})\big\},
\end{aligned}
\end{equation}
where $J_{k+1}(I_{k+1})$ is given by

\begin{equation}
J_{k+1}(I_{k+1})=\max_{P_{k+1},o_{k+1},\cdots,P_N,o_N}\left\{\Pi_{N+1}+\sum_{j=k+1}^N\Pi_j\right\}.
\end{equation}

\vfill

\end{document}